\documentclass[%
reprint,
twocolumn,
amsmath,amssymb,
aip,pop,
nofootinbib,
longbibliography, 
floatfix,
]{revtex4-2}
\usepackage[utf8]{inputenc}
\usepackage{lipsum}
\usepackage{graphicx}
\usepackage{dcolumn}
\usepackage{bm}
\usepackage{amssymb}
\usepackage{amsmath}
\usepackage{textcomp}
\usepackage{color}
\usepackage{mathtools}
\usepackage{units}
\usepackage{appendix}
\usepackage{microtype}

\begin{document}

\title{Laser ion acceleration from concave targets by subpicosecond pulses}

\author{K. V. Lezhnin}
\email{klezhnin@pppl.gov}
\affiliation{Princeton Plasma Physics Laboratory, 100 Stellarator Rd, Princeton, NJ 08540, USA\looseness=-1}
\author{V. Ospina-Bohórquez}
\affiliation{Focused Energy Inc., 600 Center Ridge Dr, \#100, Austin, TX 78753, USA\looseness=-1}
\author{J. Griff-McMahon}
\affiliation{Princeton Plasma Physics Laboratory, 100 Stellarator Rd, Princeton, NJ 08540, USA\looseness=-1}
\affiliation{Department of Astrophysical Sciences, Princeton University, Princeton, NJ 08544, USA\looseness=-1}
\author{K. Bhutwala}
\affiliation{Princeton Plasma Physics Laboratory, 100 Stellarator Rd, Princeton, NJ 08540, USA\looseness=-1}
\author{R. Nedbailo}
\affiliation{University of Texas at Austin, Austin, TX 78712, USA\looseness=-1}
\author{R. Davis}
\affiliation{University of York, Heslington, York, YO10 5DD, UK\looseness=-1}
\author{\\X. Vaisseau}
\affiliation{Focused Energy Inc., 600 Center Ridge Dr, \#100, Austin, TX 78753, USA\looseness=-1}
\author{I. D. Kaganovich}
\affiliation{Princeton Plasma Physics Laboratory, 100 Stellarator Rd, Princeton, NJ 08540, USA\looseness=-1}
\author{S. Malko}
\affiliation{Princeton Plasma Physics Laboratory, 100 Stellarator Rd, Princeton, NJ 08540, USA\looseness=-1}

\date{\today}

\begin{abstract}
Laser-driven proton acceleration provides a powerful route for generating ultrashort, high-charge proton beams. Many applications, including secondary neutron sources and inertial fusion, benefit from tight proton beam focusing. Concave targets offer a robust solution, yet the scaling of proton focusing with laser and target parameters remains poorly understood. Here, we present a numerical study of laser-driven proton acceleration and focusing from hemispherical targets using the fully kinetic, relativistic Particle-In-Cell code EPOCH. We focus on the sub-picosecond laser-pulse regime (duration $\lesssim 10^2$ fs), centering on the laser parameters of our recent experiment at the CSU ALEPH laser facility. We investigate the proton acceleration mechanisms, characterize proton focusing, and assess how focal spot parameters scale with laser and target parameters. We identify Target Normal Sheath Acceleration as the dominant mechanism, supplemented by a secondary post-acceleration stage near the geometrical center of the hemisphere. We demonstrate that both the proton focal spot size and focal plane position scale approximately linearly with the hemisphere radius, with the focal plane consistently located downstream of the geometrical center. The opening angle of the concave target mainly affects the proton beam waist. Energy-dependent proton focusing is interpreted as a consequence of the evolving curvature of the accelerating structure, which departs from the target curvature. Evidence for self-similar proton focusing is found in the regime of nearly uniform target irradiation.
\end{abstract}

\maketitle

\section{Introduction}

Laser-driven proton acceleration is a promising route to producing ultrashort, high-brilliance fast proton beams that are not readily available from conventional accelerators\cite{macchi2013}. It has a wide range of applications\cite{daido2012,schreiber2016}, including hadron therapy\cite{bulanov2002,bulanov2014}, proton imaging\cite{borghesi2001,schaeffer2023}, secondary neutron sources\cite{alvarez2014,yogo2023}, and inertial fusion\cite{roth2001}. Most applications would benefit from controlled proton beam focusing. Concave targets offer a robust and widely explored approach to focusing laser-driven protons\cite{wilks2001,ruhl2001,patel2003}, though other concepts have also been proposed\cite{toncian2006,schollmeier2008}. In particular, concave targets embedded in shielding cones are a primary target design considered for the proton fast ignition (pFI) concept of inertial fusion\cite{key2006}. In this approach, energetic protons generated by an ultraintense short pulse laser propagate into the compressed fuel and deposit their energy at the Bragg peak, enabling localized and efficient heating of the dense fuel core. Achieving ignition-relevant conditions requires proton beams with $\sim$20 kJ of energy focused to a $\sim$40 micron spot over a $\sim$10 ps timescale, with proton energies of a few MeV\cite{ditmire2023}. Despite decades of theoretical and experimental progress, current proton beam parameters remain well below these requirements. Bridging this gap has motivated a series of scaled experiments across major laser facilities worldwide\cite{offermann2011,kar2011,chen2012,bartal2012,mcguffey2020,morace2022,shi2022,griffmcmahon2026}.

The concept of proton focusing using concave targets was first introduced in Refs.\cite{wilks2001,ruhl2001}, where kinetic simulations predicted that protons accelerated via Target Normal Sheath Acceleration (TNSA) converge near the geometrical center of a hemisphere. However, these studies relied on strongly downscaled targets due to computational limitations at the time. While such scaling preserves qualitative behavior, it can significantly alter quantitative properties of the focused proton beam. In particular, in the TNSA regime, the proton acceleration scale is determined by the Debye length associated with hot electrons generated at the laser-irradiated surface, whereas focusing is set by the rear surface curvature. As a result, scaling down the target geometry can modify the relationship between acceleration and focusing processes.

Several experimental platforms have investigated laser-driven proton acceleration and focusing. The pioneering work by Patel et al.\cite{patel2003} demonstrated focusing using hemispherical targets at the Janus laser (10 J, 100 fs) by comparing isochoric proton heating of a sample with flat and hemispherical targets. Subsequent experiments\cite{snavely2007} performed XUV imaging of proton-heated Al samples with the Gekko laser (170 J, 700 fs). Later experiments at the Trident (80~J, 600~fs) and OMEGA EP (1~kJ, 10~ps)\cite{offermann2011,bartal2012,foord2012} lasers employed mesh radiography techniques to characterize proton focusing from partial and full hemispherical targets. The inferred focal positions showed varied behavior, with focusing observed either outside\cite{offermann2011} or inside\cite{bartal2012} the radius of curvature. The focal spot sizes approached the pFI-relevant conditions, and the results were similar for partial and full hemispheres. Hybrid PIC simulations with the LSP code reproduced the measured proton focal location. However, the origin of the offset between the proton focal plane and the geometrical center of the hemisphere, as well as the relationship between the actual (i.e., plane of peak proton beam compression) and measured proton focal planes, were not fully explained. In Refs.\cite{kar2011,chen2012}, hemicylinder targets were used to study proton focusing at the Vulcan (250 J, 750 fs) and LULI (1 J, 320 fs) laser facilities. Using double-mesh radiography, Ref.\cite{kar2011} characterized proton focusing by comparing proton beam properties in the focusing and non-focusing planes. The measured proton focal location and size were found to be energy-dependent. While the source sizes scaled similarly to the findings of Ref.\cite{bartal2012}, the inferred proton focal location was outside the curvature radius for all proton energies. Ref.\cite{chen2012} reported proton focusing around or past the hemicylinder center using the proton deflectometry diagnostic. Together, these results confirm the feasibility of proton focusing from curved targets, but also highlight persistent uncertainties in the relationship between measured and actual proton source properties and their dependence on laser and target parameters.

Multiple analytical and numerical approaches have been developed to explain these observations. Self-similar plasma expansion models\cite{dorozhkina1998,bellei2012} capture some qualitative features of proton beam evolution, but the underlying assumptions can limit their applicability under certain experimental conditions. An alternative model based on 2D radial isothermal plasma expansion\cite{offermann2011} reproduces certain trends but requires empirical adjustments to match experimental data, underscoring the need for first-principles simulations.

A significant effort to simulate proton focusing from hemispherical targets of realistic sizes was carried out in support of the experimental campaign reported in Refs.\cite{offermann2011,bartal2012,foord2012}. The authors used a hybrid particle-in-cell (PIC) code LSP, in which an electron beam was injected to mimic laser interaction with the front side of the target, after which the plasma was evolved self-consistently using fluid equations for the target material and kinetic descriptions of hot electrons and accelerated protons. Protons were found to focus upstream of the geometrical center of the hemisphere (i.e., inside of the curvature radius), and the effects of a shielding cone structure, hemisphere illumination, and energy-dependent focusing were identified. In Ref.~\cite{qiao2013}, hybrid Particle-in-cell (PIC) simulations with a laser–plasma interaction (LPI) module were performed to assess proton focusing from hemispherical targets more rigorously. By considering different target configurations (flat, hemispherical, and hemisphere-cone), the authors compared proton focusing characteristics and identified features of proton focusing associated with different electric field structures.

Fully kinetic simulations of laser-driven proton focusing by concave targets have also been reported. Ref.\cite{carrie2011} considered sub-micron thick curved targets and the laser parameters of Ref.\cite{patel2003}, finding proton focusing at or downstream of the geometrical center of the hemisphere (i.e., outside of the curvature radius). Simulations of a short laser pulse interacting with a downscaled (10~$\mu$m radius) hemispherical target with a wire array on the front side \cite{jiang2024} demonstrated the beneficial role of front side structures in enhancing conversion efficiency (CE), as well as the role of multi-pulse irradiation in improving proton focusing. Ref.\cite{kim2025} studied a novel mechanism of proton acceleration and focusing for ultraintense lasers and $\mu$m-scale targets with convex target front. More recently, Ref.\cite{kemp2024} used fully kinetic simulations to address laser-to-proton CE scaling with laser and target parameters, focusing primarily on the picosecond pulse duration regime. The role of the concave rear surface and a finite laser spot was also discussed in the context of CE into protons propagating through a specific volume behind the target, used as a proxy for the pre-imploded target in pFI. The role of self-generated magnetic fields in proton focusing in the picosecond pulse regime was identified in Ref.\cite{kemp2025}. Proton focal scaling with laser and concave target parameters was recently reported in the picosecond pulse regime \cite{higginson2026}. While these studies provide important insights into proton acceleration and focusing, systematic investigations of proton focusing scaling with concave target geometry as well as the connection between simulation predictions and experimentally inferred proton focal properties remain limited. 

To address these limitations, we present a systematic fully kinetic study of laser-driven proton acceleration and focusing from hemispherical targets, motivated by ongoing experiments\cite{griffmcmahon2026} at the Colorado State University (CSU) ALEPH laser facility (20 J, 40 fs). Our goal is to identify robust scaling trends in proton acceleration and focusing relevant to short-pulse ($\lesssim 10^2$ fs) laser experiments. We investigate a wide range of hemisphere radii, from 20~$\mu$m, typical of early simulation studies, to 120~$\mu$m, corresponding to the hemisphere sizes used in the CSU experiments. We further examine the influence of the hemisphere opening angle, in light of experimental observations suggesting little difference between full and partial hemispheres~\cite{bartal2012} and recent hybrid simulations that predict stronger sensitivity to this parameter~\cite{higginson2026}. Finally, we investigate the energy dependence of proton focusing and quantify how focusing properties scale with laser intensity, spot size, and duration.

This paper is organized as follows. Section~\ref{sec:theory} briefly reviews the theoretical aspects of the TNSA scheme\cite{wilks2001}. Section~\ref{sec:setup} describes the simulation setup of EPOCH simulations. Section~\ref{sec:results} is the main section, where the simulation results are described. The paper concludes with a discussion that compares our findings with the literature, evaluates the applicability of analytical models to infer the long-term proton beam evolution, and examines the impact of simulation dimensionality, contamination layer parameters, finite laser contrast, and laser pointing stability.

\section{Theoretical background}\label{sec:theory}

The most robust laser ion acceleration mechanism known to date is Target Normal Sheath Acceleration (TNSA)\cite{wilks2001}. It manifests when an intense laser pulse interacts with the front surface of a relatively thick target, generating fast electrons that traverse the target and emerge from its rear side, where they ionize the material and accelerate ions via the electrostatic sheath field. The hot electron temperature is typically assumed to follow the ponderomotive scaling (see Ref.\cite{rusby2024} for a more general discussion of electron temperature scaling):

\begin{equation}
    T_{\rm e,h} = m_ec^2 \left(\sqrt{1+a_0^2/2}-1\right).
    \label{eq:Tepond}
\end{equation}

\noindent Here, $m_e$ is the electron mass, $c$ is the speed of light in vacuum, and $a_0 = eE/m_e\omega_0c = 0.85 \sqrt{I/10^{18}\rm W/cm^2}\cdot (\lambda/1 \rm \mu m)$ is the dimensionless vector potential of the laser wave. $e$ is the elementary charge, $E$ is the amplitude of the electric field in the laser wave, $\omega_0$ is the angular frequency of the laser radiation, $I$ is the laser wave intensity, and $\lambda$ is the laser wavelength. The accelerating field may be roughly estimated as:
\begin{equation}
    E_{\rm TNSA,0} \approx \frac{k_B T_{\rm e,h}}{e \lambda_D} = \sqrt{4\pi n_{\rm e,h}k_B T_{\rm e,h}},
    \label{eq:Etnsa0}
\end{equation}

\noindent where $k_B$ is the Boltzmann constant and $\lambda_D =\sqrt{k_BT_{\rm e,h}/4\pi n_{\rm e,h}e^2}$ is the Debye length for the hot electron population. This electric field drops off with time as\cite{mora2003}:

\begin{equation}
    E_{\rm TNSA} \approx E_{\rm TNSA,0} \cdot \frac{2}{\sqrt{2e+\omega_{\rm pi}^2t^2}}.
    \label{eq:TNSAt}
\end{equation}

\noindent Here, $\omega_{\rm pi}=\sqrt{4\pi Z n_e e^2/m_i}$ is the ion plasma frequency. The cutoff energy is predicted to be proportional to the hot electron temperature, $\mathcal{E}_{\rm p} \propto T_{\rm e,h}$, and is roughly ${\rm few}\times T_{\rm e,h}$\cite{fuchs2006}. The efficiency of the TNSA mechanism therefore depends on the ability to create large $T_{\rm e,h}$ and $n_{\rm e,h}$, which is crucial for reaching the high CE demanded by pFI.

\begin{figure}
    \centering
    \includegraphics[width=\linewidth]{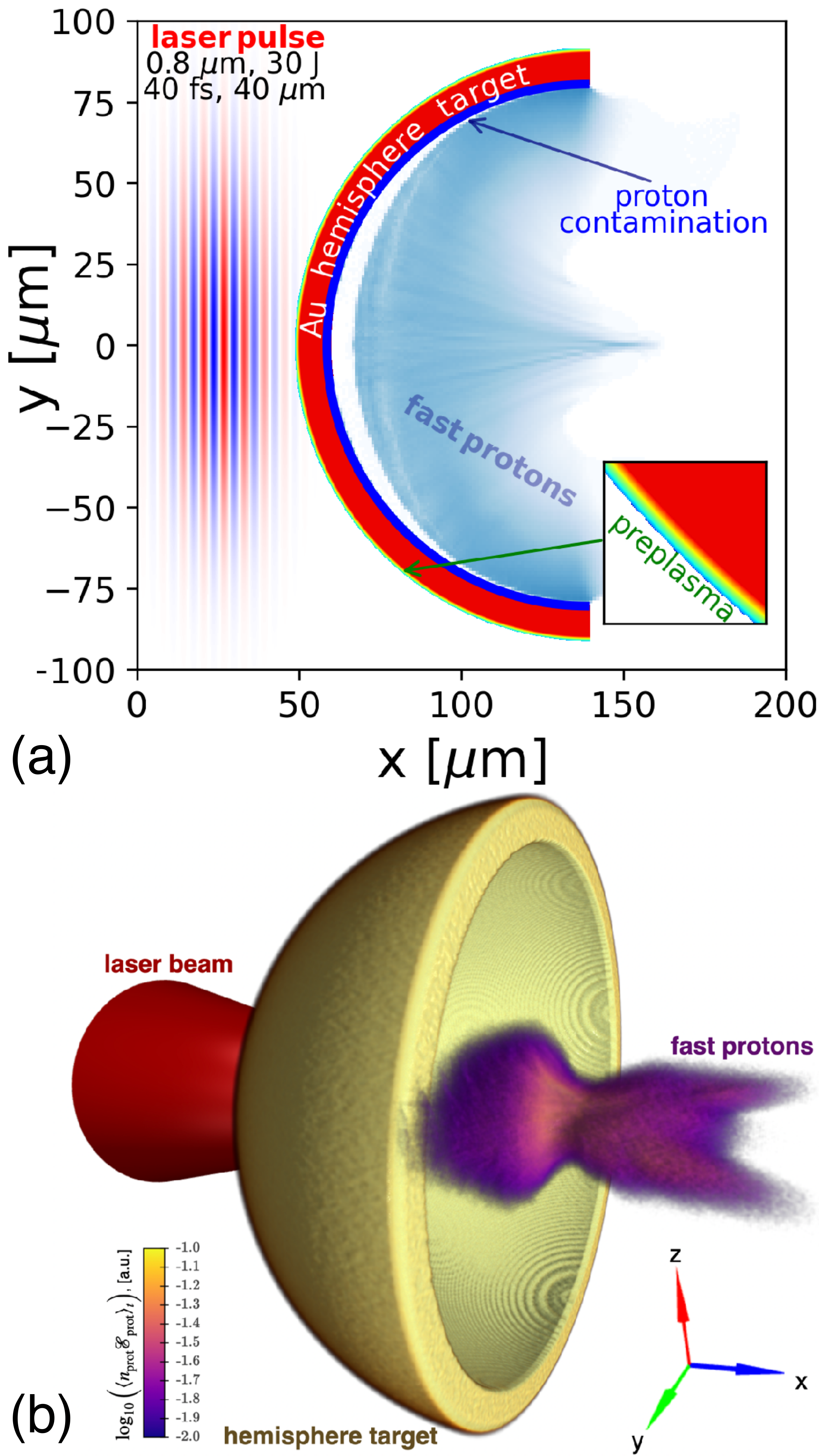}
    \caption{(a) Sketch of the 2D EPOCH simulation setup. A laser pulse enters from the left boundary and propagates along the $y=0$ axis, interacting with a gold hemispherical foil and accelerating protons from the contamination layer on the rear side of the target. A zoom-in of the front target surface highlights the presence of preplasma on the laser side of the target. (b) Illustration of 3D simulation result showing protons focused near the geometrical center of the hemispherical target.}
    \label{fig:setup}
\end{figure}

\begin{figure*}
    \centering
    \includegraphics[width=\linewidth]{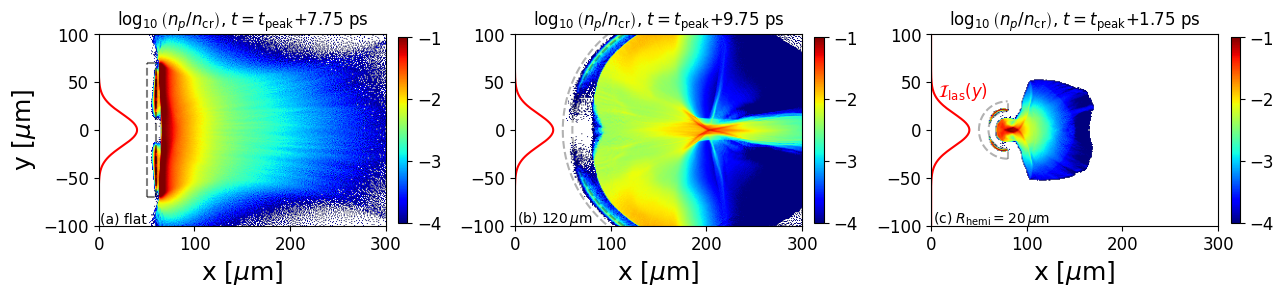}
    \caption{Proton density distributions in laser-driven proton acceleration from concave targets. (a) $R_{\rm hemi}=\infty$ (flat), (b) $R_{\rm hemi}=120\, \rm \mu m$, and (c) $R_{\rm hemi}=20\, \rm  \mu m$. Dashed lines indicate the initial target location, and red lines show the transverse laser intensity profile.}
    \label{fig:flat_vs_120_vs_20}
\end{figure*}

\section{Simulation setup}\label{sec:setup}
PIC simulations with the code EPOCH\cite{arber2015} are performed to investigate proton acceleration and focusing. Figure~\ref{fig:setup} depicts a typical simulation setup. We conduct two-dimensional (2D) Cartesian grid simulations of laser-driven proton acceleration from hemispherical targets. The box size depends on the hemisphere size: for hemisphere radius $R_{\rm hemi}$ and target thickness $d_{\rm hemi}$, we choose a box length $L_x = 50+ 3R_{\rm hemi}+d_{\rm hemi}$ and box width $L_y = 2.4 R_{\rm hemi}$ (all quantities are in microns). Target thickness is $d_{\rm hemi}=10\, \rm \mu m$, and the hemisphere radius is varied: $R_{\rm hemi} = 20,\,40,\,80,\,120\,\rm \mu m$. The grid resolution is $60$ grid cells per micron, and outflow boundary conditions are applied in all directions. The left boundary is also used to emit the laser pulse. Current smoothing\cite{buneman1993} is applied to reduce numerical self-heating of the cold, dense target. The hemispherical target center is located at $x_{\rm hemi,\, center}=50+R_{\rm hemi}+d_{\rm hemi}$. 

The target is comprised of $n_{e,\rm max}=40\,n_{\rm cr}$ gold target ($n_{\rm cr} = 1.1\cdot 10^{21}/(\lambda[\rm \mu m])^2 \, cm^{-3}$ is the critical density for a laser wavelength $\lambda$), with gold ions assigned a fixed ionization state of 11. We add a $l_{\rm prp}=$0.25$\,\mu$m exponential preplasma on the laser side of the target ($n_e = n_{e,\rm max} \exp{[-(r-R_{\rm hemi}-d_{\rm hemi})^2/l_{\rm prp}^2]}$, $R_{\rm hemi}+d_{\rm hemi} \leq r\leq R_{\rm hemi}+d_{\rm hemi}+\delta_{\rm prp}$, $\delta_{\rm prp}=1.5\,\rm \mu m$, and $r$ is measured from the hemisphere center). Protons are deposited on the rear side of the target with peak density $n_{e,\rm max}$ and exponential falloff with an e-folding scale length of $10$~nm and layer thickness of 50~nm. The choice of gold ionization level, as well as the target, preplasma, and contamination layer parameters, is inherited from prior numerical studies of laser-driven proton acceleration from gold targets under comparable laser conditions\cite{allen2004}. Protons and electrons are each resolved with 100 particles per cell, while gold ions are resolved by 9 particles per cell. Flat target simulations use the same target thickness and density, with the front side of the flat target located at $x=50\,\rm \mu m$. 

The primary laser parameters correspond to the CSU ALEPH laser\cite{griffmcmahon2026}: laser wavelength $\lambda = 0.8 \, \mu \rm m$, peak intensity $\mathcal{I}_{\rm max}=2.6 \cdot 10^{19}\, \rm W/cm^2$, beam width (FWHM) $40\, \mu \rm m$, pulse duration (FWHM) $40$ fs, with a Gaussian distribution of laser intensity. To study the role of the transverse hemisphere extent (opening angle), we conducted simulations in which the transverse size of the target was $30\%,\, 50\%,\, 70 \%,\,\rm{and}\ 100\%$ of the full width: $2(R_{\rm hemi}+d_{\rm hemi})$. Auxiliary simulations examining the role of a flat target holder with an embedded hemisphere showed no significant effect on proton energy spectra or focusing properties.

We also conducted convergence studies for the quasi-1D simulations with respect to box size ($L_x=400,1000\,\mu$m), grid resolution ($N_{\rm grid}=60, 100, 150$ per micron), number of particles per cell ($N_{\rm ppc}=50,100,200$), current smoothing on/off, and peak target density ($n_{e,\rm max}/n_{\rm cr}=40, 80, 120$), finding reasonably good convergence of the large-scale features of the expanding plasma profiles. Outflow boundary conditions were applied in the longer box direction, while periodic boundary conditions were used in the shorter one. Based on this study, we found that $L_x=400\,\mu$m, $N_{\rm grid}=60$, $N_{\rm ppc}=100$, $n_{e,\rm max}/n_{\rm cr}=40$, with current smoothing enabled, is sufficient to produce converged results.

To assess the effects of reduced dimensionality, we conducted two closely matched simulations in 2D and 3D for $R_{\rm hemi}=20~\mu \rm m$. Figure~\ref{fig:setup}b sketches out the results of a 3D simulation. Although the simulations exhibit the expected dimensionality effects, the predicted proton focusing characteristics are in qualitative agreement, thereby justifying the use of the 2D approach throughout the rest of the manuscript. See Appendix~\ref{sec:3D} for details of the comparison between the 2D and 3D simulations.

\section{Results}\label{sec:results}
In this section, we discuss the proton acceleration mechanism, the properties of the electron–proton plasma at peak compression, the origin of energy-dependent proton focusing, and the scaling of proton focusing with laser and target parameters.

Figure~\ref{fig:flat_vs_120_vs_20} depicts typical proton density distributions in three representative simulations: (a) a flat target ($R_{\rm hemi} = \infty$), (b) $R_{\rm hemi} = 120 \rm \, \mu m$, and (c) $R_{\rm hemi}=20 \, \rm \mu m$. Here, the flat target is the baseline case; $R_{\rm hemi} = 120 \rm \, \mu m$ is representative of the smallest size of hemispherical targets available experimentally, yet still trackable with our 2D PIC simulations; $R_{\rm hemi}=20 \, \rm \mu m$ is the smallest hemisphere size we simulate, which exaggerates the role of the concave target surfaces, making their effects easier to diagnose. It is evident that the concave rear surface shapes proton density distributions by causing them to focus around the center of the hemispherical target. While focusing is evident in Figs.~\ref{fig:flat_vs_120_vs_20}b,c, defocusing of the compressed electron-proton plasma is also observed further downstream. The degree of both focusing and defocusing clearly depends on $R_{\rm hemi}$.

\subsection{Acceleration mechanism}

To get a better understanding of the accelerating fields in these cases, we compare ion acceleration properties along the laser axis ($|y|\leq 1\, \rm \mu m$). While the dynamics of the electron-proton plasma is inherently two-dimensional even in simulations with a flat target (note defocusing fringes around $x=150 \rm \, \mu m$ in Fig.~\ref{fig:flat_vs_120_vs_20}a), this comparison helps highlight the effects that come into play for a finite hemisphere radius.

\begin{figure}
    \centering
    \includegraphics[width=\linewidth]{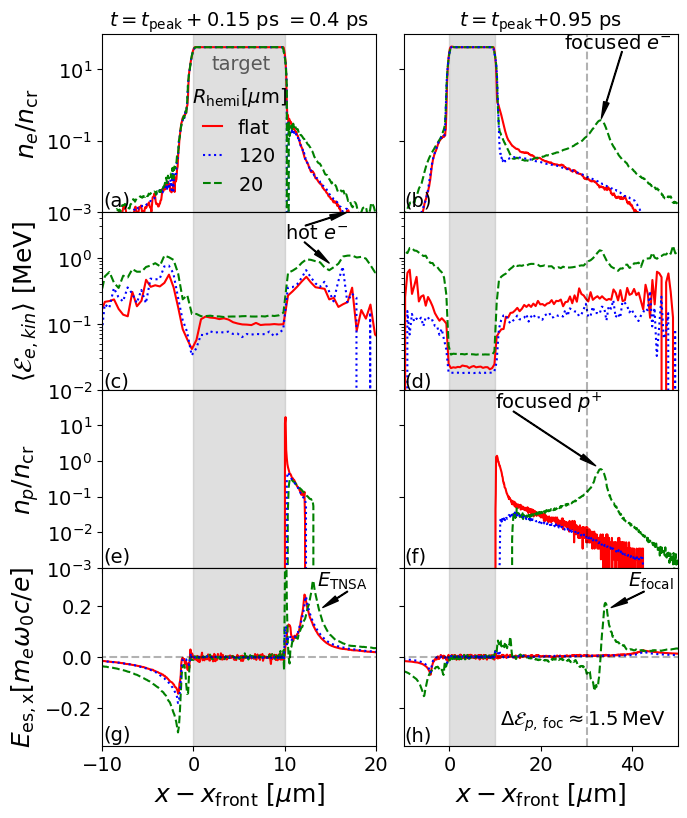}
    \caption{Acceleration mechanism in quasi-1D (along the laser axis) for $R=\infty$ (flat, red), $120$ (dotted blue line), and $20~\mu$m (dashed green line) hemispheres. (a,b) Electron densities, (c,d) electron temperatures, (e,f) proton densities, and (g,h) longitudinal electrostatic fields at $t=t_{\rm peak}+0.15~\rm ps =0.4$~ps (a,c,e,g) and $t=t_{\rm peak}+0.95$~ps (b,d,f,h). Vertical semitransparent dashed line denotes $R_{\rm hemi}=20\,\mu$m center. TNSA-like acceleration, a post-acceleration stage, and plasma focusing for $R_{\rm hemi}=20~\mu$m are evident.}
    \label{fig:accmech1D}
\end{figure}

Figure~\ref{fig:accmech1D} compares longitudinal plasma dynamics in the three representative simulations. Here, we show (a-b) electron density, (c-d) mean electron energy (hot electron temperature), (e-f) proton density, and (g-h) the $x$ component of the electrostatic field. The left column shows a snapshot at 0.15~ps after the laser peak arrives at the front side of the target (denoted as $t_{\rm peak}=0.25\,$ps) and the right column shows a snapshot at the time of peak proton beam focusing in the $R_{\rm hemi}=20~\mu \rm m$ case at $t_{\rm peak}+0.95~\rm ps$. Red lines correspond to the flat target, blue dotted lines to $R_{\rm hemi}=120~\mu \rm m$ hemisphere, and green dashed lines to $R_{\rm hemi}=20~\mu \rm m$.

At early time (left column), all three simulations demonstrate qualitatively similar on-axis behavior. A hot electron population with nearly exponential profiles is present on both sides of the targets (Fig.~\ref{fig:accmech1D}a). The mean electron energy approaches the ponderomotive prediction $T_{e,h} \approx 0.8~ \rm MeV$ (Fig.~\ref{fig:accmech1D}c \& Eq.~\ref{eq:Tepond}). Similarly, protons begin to develop an exponential density profile (Fig.~\ref{fig:accmech1D}e), and the sheath field associated with the ion front is observed (Fig.~\ref{fig:accmech1D}g). 

Notably, the electron temperature is higher in the $R_{\rm hemi}=20~\mu \rm m$ case. We conducted two auxiliary simulations where we considered (1) a target with a flat front surface and a curved rear surface ($R_{\rm curv}=20~\mu \rm m$) and (2) a target with a curved front surface and a flat rear surface. Simulation (1) predicted a $T_{e,h}$ similar to that of the flat target, while simulation (2) was close to the $R_{\rm hemi}=20~\mu \rm m$ result. We conclude that the curvature of the front surface contributes to the higher $T_{e,h}$ values. By inspecting the fields normal to the front target surface, we found that a curved front surface with $R_{\rm curv}=20~\mu \rm m$ produces fields that are $\approx 2.3$ times stronger than those for a flat front surface, explaining the increased $T_{e,h}$. One possible interpretation is that for the targets with $R_{\rm hemi}\sim w_{\rm las}$, part of the pulse experiences oblique incidence, which is known to enhance laser-to-electron energy conversion efficiency \cite{ping2008absorption}.

The later time evolution highlights how ion acceleration differs for a small hemispherical target. At $t=t_{\rm peak}+0.95\,\rm ps$, the system approaches the proton focusing time for the $R_{\rm hemi}=20~\mu \rm m$ target, given by $t_{\rm pfoc}=R_{\rm hemi}/C_S \approx 0.1 {\rm ps}\cdot\left(R_{\rm hemi}/1\,\mu \rm m\right)\cdot \left(T_{\rm e,h}/1\rm~ MeV\right)^{-1/2}$. This expression yields $t_{\rm pfoc}=2~\rm ps$ for $R_{\rm hemi}=20~\mu \rm m$, where $C_S=\sqrt{k_BT_{e,h}/m_p}$ is the ion sound speed. Around this time, peak electron-proton density compression is observed near the hemisphere center (Fig.~\ref{fig:accmech1D}b,f) for $R_{\rm hemi}=20~\mu \rm m$. By contrast, the flat target and the $R_{\rm hemi}=120~\mu \rm m$ hemisphere remain nearly identical, since $t_{\rm pfoc}\approx 12~\rm ps$ for $R_{\rm hemi}=120~\mu \rm m$. The faster adiabatic cooling of the hot electron cloud in the flat and $R_{\rm hemi}=120~\mu \rm m$ cases also leads to a significant reduction in hot electron energies compared with the $R_{\rm hemi}=20~\mu \rm m$ case (Fig.~\ref{fig:accmech1D}d). 

The simulations reveal that plasma compression near the center of the $R_{\rm hemi}=20~\mu \rm m$ hemisphere leads to further electron energization (see small peak near the $x=85~ \mu \rm m$ in Fig.~\ref{fig:accmech1D}d), and that uncompensated proton charge at the proton focal spot leads to a build-up of electrostatic field near the focus (seen in Fig.~\ref{fig:accmech1D}h around $x=85~ \mu \rm m$). Integrating this potential along $x$, one finds that it can deliver an additional $\approx1.5~\rm MeV$ of energy to a proton traversing this region. Therefore, we conclude that while the overall ion acceleration picture is consistent with TNSA, there is a secondary post-acceleration stage that may affect the resulting proton energization.

\begin{figure}
    \centering
    \includegraphics[width=\linewidth]{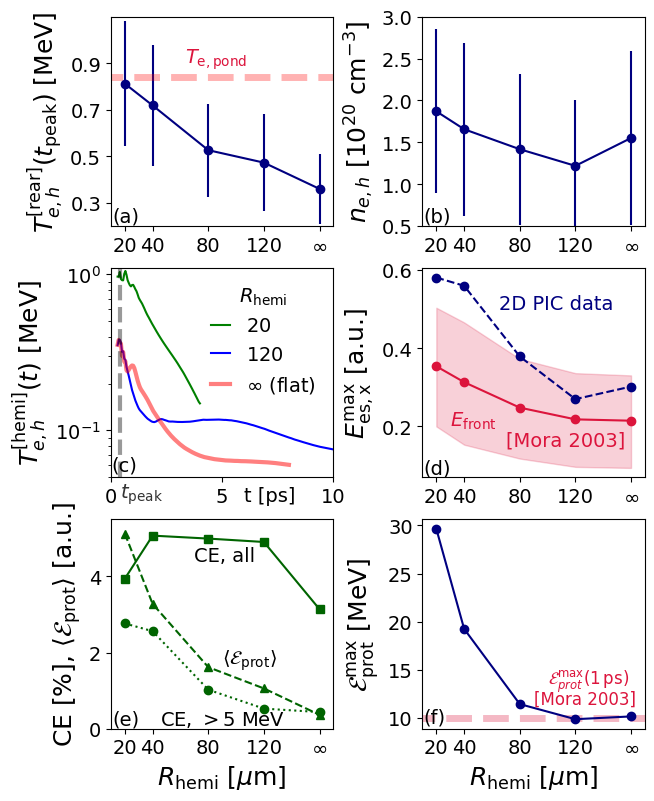}
    \caption{Dependence of laser ion acceleration on hemisphere target radius in 2D PIC simulations. (a) Hot electron temperature and (b) hot electron density at the rear side of the target at $t=t_{\rm peak}+0.15$ ps. (c) Temporal evolution of the hot electron cloud temperature. (d) Peak accelerating field as a function of $R_{\rm hemi}$. (e) Laser-to-proton conversion efficiency (all protons, fast protons only, and normalized by the total number of protons). (f) Maximum proton energy at $t=1.25$ ps for $R_{\rm hemi}=20,\,40,\,80,\,120,\,\infty~\mu$m. Theoretical estimates for the hot electron temperature, peak accelerating field, and proton cutoff energy (based on Refs.~\cite{wilks2001,mora2003}) are also shown.}
    \label{fig:TNSAscaling}
\end{figure}

\begin{figure}
    \centering
    \includegraphics[width=\linewidth]{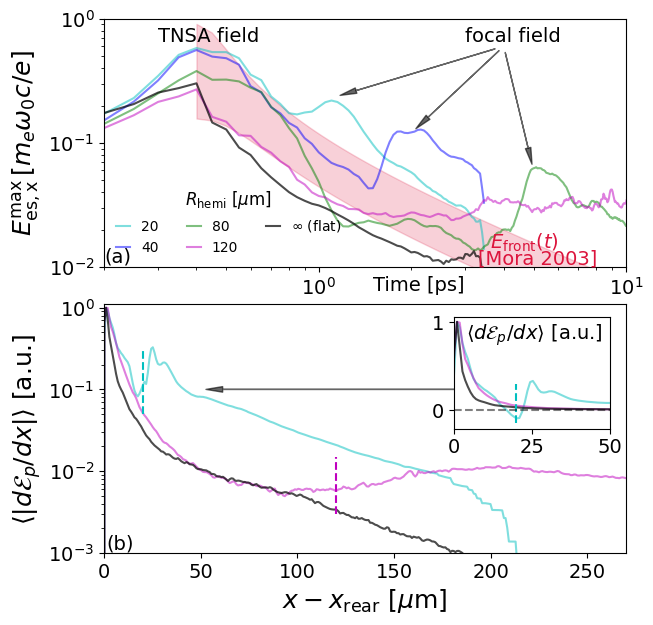}
    \caption{Role of the sheath field and secondary acceleration in hemispherical targets. (a) Time evolution of the peak electrostatic field along $x$ in 2D PIC simulations for varied $R_{\rm hemi}$. The red band shows analytical field estimates from Eq.~\ref{eq:TNSAt}. (b) Absolute and signed (inset) proton energy gain per unit length along the laser axis for $R_{\rm hemi}=20,\,120,\, \infty~\rm \mu$m. Vertical dashed lines depict the location of the geometrical center of the respective hemisphere. The contribution of secondary acceleration near the geometrical center is highlighted.}
    \label{fig:accfields1D}
\end{figure}

\subsection{The role of the hemisphere radius}

To analyze whether the TNSA mechanism is affected by the curved front and rear sides of the target, we calculate acceleration-related quantities at the rear side of the target near the laser axis at $t=t_{\rm peak}+0.15~\rm ps$. Figure~\ref{fig:TNSAscaling} summarizes the hot electron temperature (a) and density (b) at the target rear (measured in a $1~\mu \rm m\times 1~\mu \rm m$ square located at $x=x_{\rm rear}+1~\mu \rm m$). It also shows average electron temperature inside the hemisphere (c), along with the peak electric field (d), laser-to-proton conversion efficiency (e), and peak proton energy (f) as functions of hemisphere radius. Error bars in (a,b) were obtained by varying the collection time by $\pm 0.1~\rm ps$.

The hot electron density is essentially independent of the hemisphere radius, as its spatial distribution is not strongly affected by curvature. The hot electron temperature, while roughly consistent with the ponderomotive prediction, drops by about a factor of two for larger $R_{\rm hemi}$. As discussed above, this may be attributed to enhanced laser-to-electron absorption on the curved front side of the target. Accordingly, the peak accelerating electric field decreases for larger $R_{\rm hemi}$. Figure \ref{fig:TNSAscaling}d estimates the electric field from Eq.~\ref{eq:Etnsa0} using the measured hot electron density and temperature (red line, with the shaded red region corresponding to the error bars in Figs.~\ref{fig:TNSAscaling}(a,b)) and shows fair agreement with the accelerating field observed in the simulations.

Fig.~\ref{fig:TNSAscaling}e depicts laser-to-proton energy conversion efficiency (CE) using several different metrics. The solid line calculates CE based on the entire proton spectrum, the dotted line calculates CE for protons with final energy $>$5 MeV (a common threshold for proton energies relevant to fuel heating in pFI\cite{key2006}), and the dashed line plots the average energy per proton in arbitrary units. We calculate conventional CE to be $3-5\%$ for all radii, with no clear trend with $R_{\rm hemi}$. Accounting for the different total numbers of protons in each simulation (we deposit 50 nm proton layer on the rear side of the hemisphere/flat target, so the total proton number depends on the target size and type), we find a clear decrease in the average energy per proton. A similar trend is observed when we recalculate CE for protons with final energy $>5$ MeV. 

The maximum proton energy at $t=t_{\rm peak}+1\,\rm ps$ saturates at large $R_{\rm hemi}$. Remarkably, small hemispheres produce much larger proton energies due to both higher initial $T_{e,h}^{\rm rear}$ and slower adiabatic cooling of the hot electron cloud. The hot electron temperature evolution inside the hemisphere (Fig.~\ref{fig:TNSAscaling}c; for the flat target, $2\,\mu$m away from the flat rear surface) shows that during the first 4 ps, $T_{e,h}^{\rm hemi}$ for $R_{\rm hemi}=20~\mu \rm m$ stays higher than for the flat target and $R_{\rm hemi}=120~\mu \rm m$, explaining the higher proton energies. 

We therefore conclude that, for the given laser pulse parameters, smaller hemispheres accelerate fast protons more efficiently due to enhanced laser-to-electron coupling, resulting in higher initial hot electron temperatures and better hot electron retention. For larger hemispheres, the effect of target curvature on proton energization becomes negligible.

The role of the extended fields is evident from Fig.~\ref{fig:accfields1D}a. Here, we track the time evolution of the peak accelerating field on the laser axis for different $R_{\rm hemi}$. While the magnitude and temporal evolution of the electric field are generally consistent with the analytical prediction (Eq.~\ref{eq:TNSAt}), we observe a distinct secondary peak associated with the acceleration around the focal plasma. The associated proton energy gain (Fig.~\ref{fig:accfields1D}b) is sizeable for the smaller hemispheres, but becomes negligible for the $R_{\rm hemi}=120~\mu \rm m$ case. Nevertheless, this highlights the possibility of the electrostatic field build-up around the proton focal spot, which may affect focusing.

\subsection{Focal plasma properties}

\begin{figure}
    \centering
    \includegraphics[width=\linewidth]{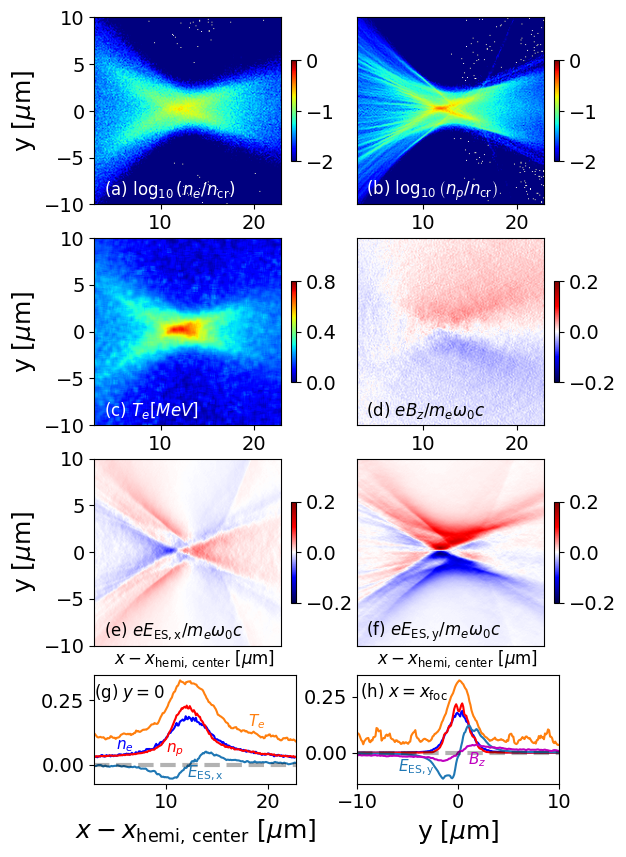}
    \caption{Properties of the focal plasma at peak compression. (a) Electron density, (b) proton density, (c) mean electron energy (``temperature''), (d) magnetic field, (e) $x$- and (f) $y$-components of the electrostatic field, (g) 1D cuts along the laser axis, and (h) 1D cuts through the location of peak plasma compression.}
    \label{fig:focalplasma}
\end{figure}

Conditions in the focal plasma are of special interest, since the main goal of using concave targets is to focus fast protons. Figure~\ref{fig:focalplasma} depicts the plasma and EM fields around the proton focal spot at the time of peak proton compression for the representative simulation with $R_{\rm hemi}=80~\mu \rm m$. First, we note that the focal spot is shifted downstream of the geometrical center of the hemisphere (note that the origin of the $x$ axis is shifted to the geometrical center of the hemisphere). The electron density is 0.1$n_{\rm cr}$, with electron temperature $T_e\approx 0.8~\rm MeV$. Notable charge separation is present (${\rm max}(n_e-n_p)/n_e\approx0.3$), and dimensionless electric fields of $E_{\rm es}^{\rm max}=0.1$ and a dipole-like magnetic field structure with $B_{\rm max}=0.03$ are observed. The longitudinal and transverse 1D cuts are shown in subplots (g) and (h). The focal plasma has an X-point-like geometry in the 2D plots, but near the peak compression point it can be fairly fitted with a Gaussian. The transverse electric field profile is defocusing for the incoming protons and may even completely deflect lower energy protons. The magnetic field is smaller by a factor of 3, and the $\bf{v\times B}$ force for non-relativistic protons (here, we have $\mathcal{E}_p \leq 30~\rm MeV$) may be neglected.

\begin{figure}
    \centering
    \includegraphics[width=\linewidth]{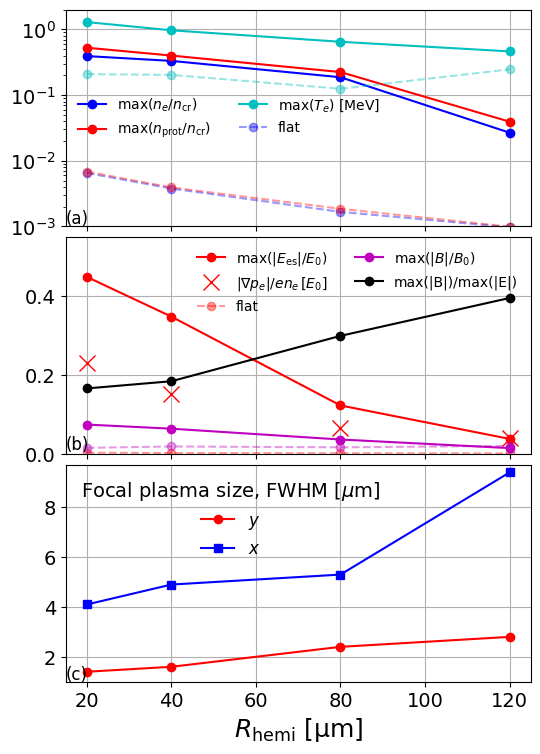}
    \caption{Properties of the focal plasma at peak compression: scaling with hemisphere radius. (a) Peak electron and proton densities, and electron temperature. (b) Peak electrostatic and magnetic fields, along with an estimate of the electric field $\nabla p_e/en_e$ and the ratio of magnetic to electric fields. (c) Focal plasma size.}
    \label{fig:focalplasmascaling}
\end{figure}

We continue by studying how the focal plasma parameters scale with hemisphere radius. Figure~\ref{fig:focalplasmascaling} summarizes the parameters discussed above, illustrated in Fig.~\ref{fig:focalplasma} for $R_{\rm hemi}=80~\rm \mu m$. Subplot (a) depicts the peak electron density (blue), proton density (red), and electron temperature (cyan). The proton density and electron temperature at the focal spot decrease with $R_{\rm hemi}$, consistent with longer dynamical times ($R_{\rm hemi}/C_S$) leading to stronger adiabatic cooling. The flat target case (dashed lines) yields substantially lower proton densities, while the electron temperatures are smaller by a factor of a few. Subplot (b) shows the peak electrostatic (red) and magnetic (magenta) fields. Both the electric and magnetic fields also fall off with $R_{\rm hemi}$. Interestingly, their ratio grows and reaches 0.4 for the largest hemispheres considered. The magnetic field is still not expected to play a dramatic role for the protons, since they are non-relativistic; we will later demonstrate this using particle tracking. To assess whether the ambipolar electric field, $\nabla p_e/en_e$, provides a reasonable estimate of the peak fields observed in our simulations, we compare it with the PIC results. We find that this estimate underpredicts the electric field for smaller $R_{\rm hemi}$, likely because it neglects contributions from uncompensated charge, but approaches the PIC values for larger $R_{\rm hemi}$. The flat target estimates are again negligible. Subplot (c) depicts FWHM measurements of the longitudinal (blue) and transverse (red) focal plasma density profiles. The proton focal spot size, estimated from Gaussian fits to 1D cuts through the focal plasma, is approximately $6\,\mu \rm m \times 2\,\mu \rm m$, with only a weak trend toward larger sizes for larger hemispheres.

In summary, hemispherical targets produce a dense, hot electron-proton plasma near the hemisphere center, with electric fields strong enough to deflect or even fully reflect protons, thereby limiting the attainable proton focal spot size.

\subsection{Proton beam waist and focal shift}

\begin{figure}
    \centering
    \includegraphics[width=\linewidth]{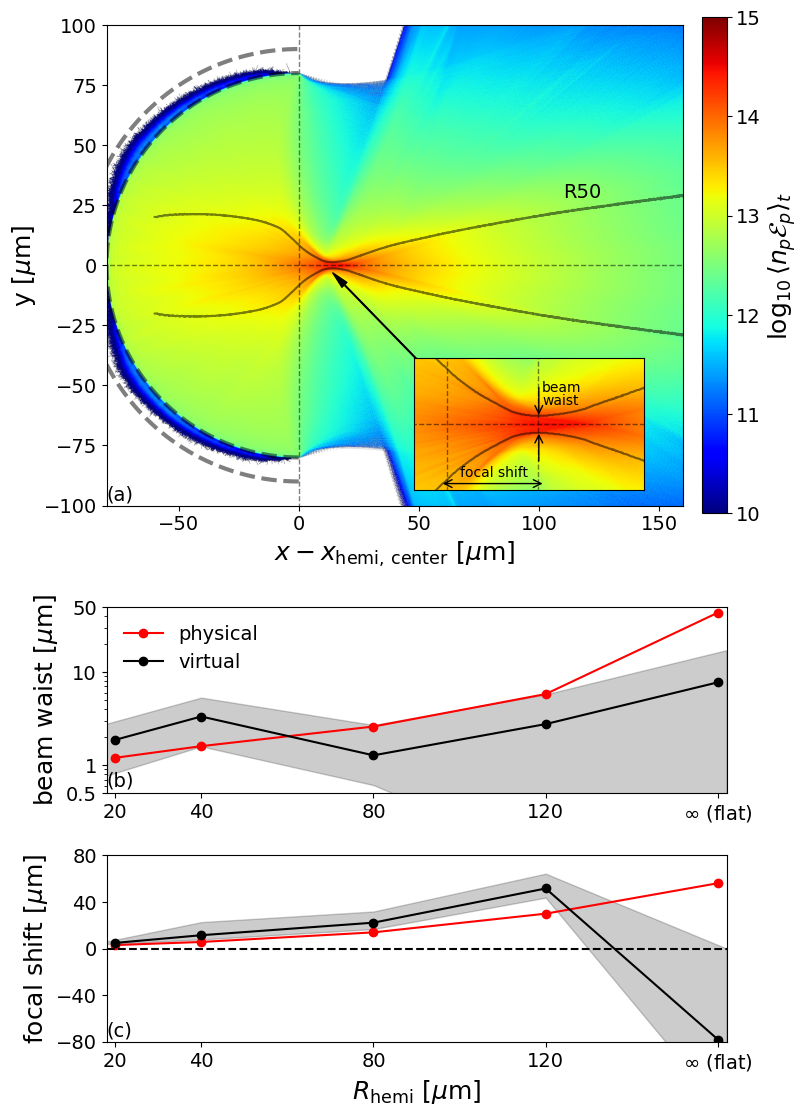}
    \caption{Scaling of proton focal characteristics with $R_{\rm hemi}$. (a) Time-integrated proton energy density in a simulation with $R_{\rm hemi}=80\,\mu \rm m$. The laser axis and the geometrical center of the hemisphere are highlighted with thin dashed lines. Thick dashed lines depict the initial location of the hemisphere target. Solid lines are R50 curves - lines enclosing 50\% of the metric of interest, calculated individually for each bin along $x$. The inset defines the proton focal plane shift and proton beam waist (focal shift and beam waist for brevity). (b) Scaling of beam waist with $R_{\rm hemi}$. Physical proton focusing (red circled line) and the inferred size of the virtual focal spot with its variance (black line and shaded grey filled region) are depicted. (c) Scaling of focal shift with $R_{\rm hemi}$.}
    \label{fig:R50waistshift}
\end{figure}

The most experimentally relevant proton focal spot parameters are the focal spot size (proton beam waist) and the location of the proton focal plane. Understanding how these scale with target parameters and predicting their values is useful for experimental design and interpretation. Ultimately, achieving control over the proton focal spot location and size is vital for future pFI-related applications. Figure~\ref{fig:R50waistshift} shows the trends in focal spot size and focal location as functions of target radius. 

To define the proton focal spot, following Ref.\cite{bellei2012}, we calculate a time-integrated proton energy density distribution, depicted on subplot (a) for $R_{\rm hemi}=80~\mu \rm m$. This diagram is then used to determine the location of the peak, with its shift from the hemisphere center defined as a focal shift, and the width, calculated as a minimum diameter containing 50$\%$ of time-integrated proton energy density, as a beam waist. To account for possible dependence of the focal parameters on their definition, we also recompute the beam waist and focal shift using time-integrated proton density, instantaneous proton density (i.e., selecting time of peak proton compression around the hemispherical center and using that snapshot to estimate the waist and shift), and, analogously, instantaneous proton energy density. All of these approaches yield similar results, therefore, only the time-integrated proton energy density calculation is shown. 

Experimentally, the proton focal characteristics are often inferred using mesh radiography\cite{kar2011,bartal2012,griffmcmahon2026}, in which a mesh is placed between the focal plane and a detector. The mesh casts a shadow of the protons onto the detector and the mesh image is used to access properties of the virtual focus, i.e. the apparent proton source under the assumption of ballistic propagation. To study the relation between the physical and virtual foci, we inferred the virtual focus from our simulations by ballistically backpropagating protons from different energy bands (all protons, the fastest protons for a given radius, 5--6 MeV, and 9--11 MeV) for a range of collection times. Next, we calculated time-integrated proton energy density maps and R50 curves in a manner similar to that used for the physical focus calculations.

We find that both the proton beam waist and the focal shift increase roughly linearly with the hemisphere radius. The proton spot size remains within a few microns, whereas flat targets yield values close to the laser FWHM. The focal shift grows with $R_{\rm hemi}$ and remains positive for all targets. We note that, for flat targets, the focal shift is calculated with respect to the target rear surface.

Virtual focal spot calculations (black line with circle markers) also suggest that the proton spot size grows with $R_{\rm hemi}$ and remains within the 1--10 $\mu$m range. The shift of the virtual focal spot remains positive for finite hemispheres, while for the flat target it shifts to the front of the target, a behavior characteristic of virtual foci in flat targets\cite{borghesi2004}. We should note that there is substantial variance in our virtual focal spot calculations depending on the definition (i.e., on the considered energy band and proton collection time; shown with grey shaded region).

Our analysis of particle trajectories suggests that, although protons beyond the focal plane can be fitted with a straight line with $R^2\geq 0.99$, the robustness of these fits degrades when the spatial window used to collect proton coordinates is varied. As the window moves away from the focal spot, a drift in the fit parameters is observed. This suggests that the proton trajectories are still evolving as they exit the computational domain, and therefore the virtual focal spot parameters may be biased. An accurate estimate of the virtual focal spot parameters thus requires simulations with a $>1$~mm-sized computational domain, which is computationally prohibitive for our current simulation approach. Even within these limitations, we find that the virtual characteristics roughly track the physical ones, with the exception of the flat foil which predicts a negative focal shift for the virtual focus.

\begin{figure}
    \centering
    \includegraphics[width=\linewidth]{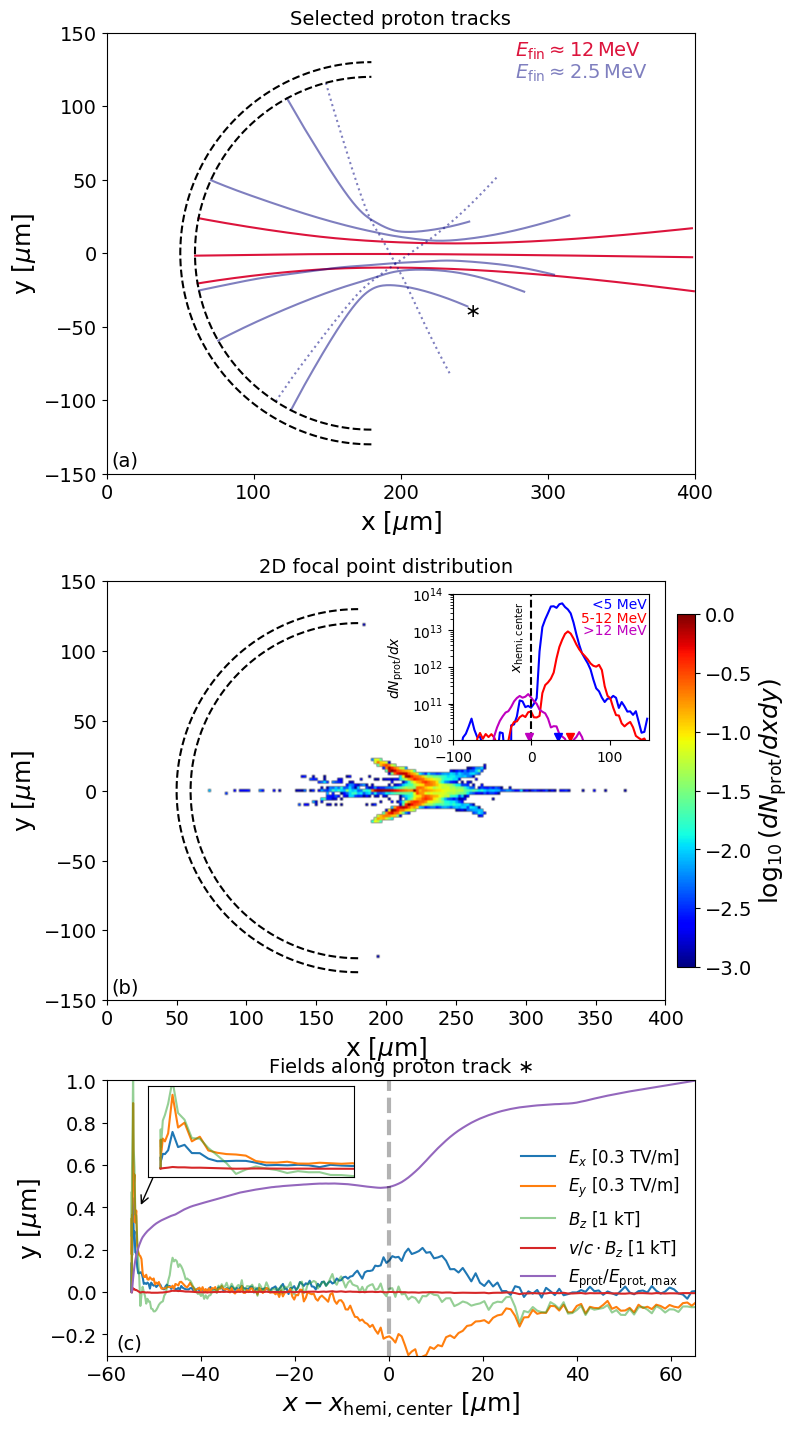}
    \caption{Proton tracking for the $R=120~\mu \rm m$ case. (a) Selected proton tracks that end up at 2.5 MeV (semitransparent navy) and 12 MeV (crimson) are shown. Dotted lines denote protons that cross the $y=0$ plane. Dashed lines depict the hemisphere location. The asterisk marks the proton trajectory for which we track electromagnetic fields along the path in subplot (c). (b) Distribution of proton focal points (see text) in 2D space. The inset shows the proton focal point distribution along $x$. Fastest ($>$12 MeV final energy, magenta), intermediate (between 5 and 12 MeV, red), slow protons ($<$5 MeV, blue), and all protons (thick grey line) are shown. Focal plane locations for these energy bands are depicted with the downward pointing triangles. (c) Electromagnetic fields along the selected proton path: $E_x$ (blue), $E_y$ (orange), $B_z$ (green), magnetic Lorentz force $|\bf{v\times B}|$ (red), and proton energy evolution over time (purple). The inset shows values early in the acceleration.}
    \label{fig:tracking}
\end{figure}

\subsection{Particle tracking and energy-dependent proton focusing}

To better understand the focal plane shift in our simulations, we perform particle tracking. During the simulation, we record all fields and proton data every 50 fs. We then downselect protons that had energies $\mathcal{E}_{\rm prot}>1~\rm MeV$ at t=7 ps and keep every 50th particle, resulting in approximately 56,000 recorded trajectories.

A few representative trajectories highlighting characteristic proton behavior are shown in Fig.~\ref{fig:tracking}a. We plot protons with final energies of 2.5 MeV (semitransparent navy) and 12 MeV (crimson). Dotted lines indicate two trajectories that cross the $y=0$ plane. Two types of proton trajectories are evident: (I) trajectories with little or no deflection from the initial direction, and (II) trajectories that are reflected before reaching the $y=0$ plane. For type (I) trajectories, we define the focusing location as the position along the $x$ axis where the particle intersects the $y=0$ plane. For type (II) trajectories, we identify the location where $p_y=0$ and record it as the focal point. Fully reflecting trajectories (type II) make up the dominant population, comprising more than 90\% of the recorded trajectories.

Fig.~\ref{fig:tracking}b depicts the distribution of the focal points in the (x,y) plane. It lies mostly downstream of the hemisphere center and is squeezed close to the $y=0$ axis, while exhibiting an X-point-like structure. By calculating the focal point distribution along the $x$ axis (see inset), we can determine the location of the proton focal plane, $x_{\rm focal}-x_{\rm hemi,\,center}\approx 35~\mu \rm m$, fairly consistent with our finding from Fig.~\ref{fig:focalplasmascaling}c (there, $x_{\rm focal}-x_{\rm hemi,\,center}\approx 29~\mu \rm m$ from the time-integrated proton energy density approach). Examining the energy dependence of the proton focal plane location, we observe a trend in which faster particles (those that ended the simulation with energies between 5 and 12 MeV; depicted in red) are focused farther downstream of the hemisphere center ($x_{\rm focal}-x_{\rm hemi,\,center}\approx 49~\mu \rm m$) than slower ones ($<5$ MeV, depicted in blue; $x_{\rm focal}-x_{\rm hemi,\,center}\approx 34~\mu \rm m$). One interesting feature is that the fastest protons ($>$12 MeV, magenta), which constitute only a small fraction of the proton population, are focused near the geometrical center of the hemisphere. This may be interpreted as follows: the fastest protons are accelerated when the curvature of the accelerating field is nearly identical to that of the target, and there is little or no plasma around the geometrical center by the time they cross it. They are therefore expected to propagate ballistically toward the geometrical center.

Throughout proton acceleration and focusing, particles may be affected by the evolving electromagnetic fields. To demonstrate the role of different field components in particle acceleration and focusing, we track the electric and magnetic fields, along with particle velocity and energy, for the selected trajectory (denoted with an asterisk in Fig.~\ref{fig:tracking}a). One may see that there are two locations along the $x$ axis where the particle is affected by the fields most strongly. 

First, at the rear target surface (see inset of Fig.~\ref{fig:tracking}c for a zoom-in on the early time field evolution), the particle is energized by strong and rapidly damping TNSA electric fields. Since its initial location is away from the laser axis, both $E_x$ and $E_y$ components are significant and reach close to 0.3 TV/m. The magnetic field on the target surface likely originates from the ``fountain effect''\cite{pukhov2001,sarri2012} and reaches a peak magnitude of 1 kT. Since the protons under consideration are non-relativistic and the dimensionless electric field is at least as large as the dimensionless magnetic field, the calculated magnetic component of the Lorentz force, $|\bf{v\times B}|$, is negligible along the entire particle path compared to $|\bf{E}|$. We also confirmed that $|\bf{v\times B}|\ll |\bf{E}|$ holds for all tracked protons, with the 90\% quantile of the $|\bf{v\times B}|/|\bf{E}|$ distribution over all trajectories being equal to 0.06.

The second location where the particle’s energy changes most strongly is near the center of the hemisphere. Here, one may see both an accelerating $E_x$ structure and a negative $E_y$ that reflects the proton back in the negative $y$ direction. We therefore conclude that magnetic fields contribute weakly to the proton dynamics, which is governed primarily by electric fields at the target surface and near the proton focal plane. We note that a recent paper\cite{kemp2025} demonstrated that magnetic fields become important for proton acceleration by picosecond laser pulses. For shorter pulses, they found that the role of the magnetic field is negligible, in agreement with our findings.

\begin{figure}
    \centering
    \includegraphics[width=\linewidth]{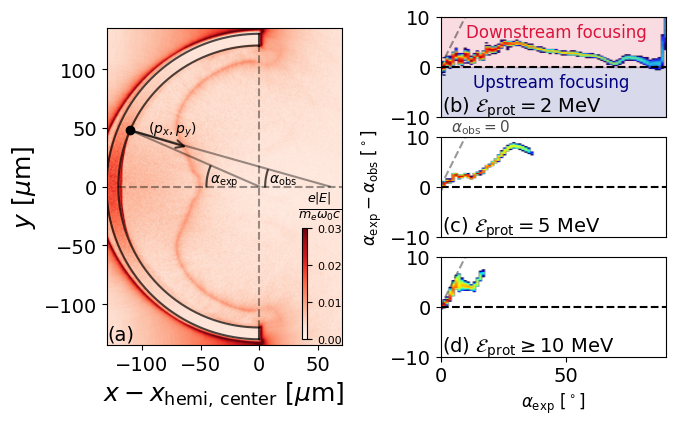}
    \caption{Proton focal shift due to the geometry of the accelerating structure. (a) Electric field amplitude distribution at $t=2~\rm ps$. Dashed lines depict the laser axis and the hemisphere center; solid black lines outline the initial location of the hemisphere target. The definitions of the expected proton focusing angle, $\alpha_{\rm exp}$, and the observed proton focusing angle, $\alpha_{\rm obs}$, are shown. Distributions of expected and observed focal angles for protons of different energies at $t=7$ ps: (b) $2$ MeV, (c) $5$ MeV, (d) $\geq 10$ MeV. Hotter colors indicate bins containing more particles. The horizontal dashed line indicates $\alpha_{\rm exp}=\alpha_{\rm obs}$ (purely geometric focusing); semi-transparent dashed lines denote proton acceleration along the x axis ($\alpha_{\rm obs}=0$).}
    \label{fig:focal_shift}
\end{figure}

To further understand the effect of the energy-dependent proton focal shift, we perform the following analysis. First, we record the initial angles at which the protons are located on the rear surface of the hemisphere. Figure~\ref{fig:focal_shift}a depicts the definition of this angle, denoted as $\alpha_{\rm exp}$. Under the assumption of geometric focusing, we would expect protons to focus exactly at the hemisphere center. Next, for all protons, we estimate the initial momentum $(p_x,p_y)$ gained at $t=2$ ps in the simulation. As is well known for the TNSA mechanism, most proton energy is gained close to the target surface, which is supported by our tracking analysis. Therefore, by estimating the angle that the proton momentum makes with the laser axis ($y=0$), we can quantify the angle at which we observe the protons focusing, assuming there is no interaction after the initial TNSA-like acceleration process. We denote this angle as $\alpha_{\rm obs}$.

Figure~\ref{fig:focal_shift}a depicts the magnitude of the electric field at $t=2$ ps. The field front is evidently not circular: the central part within the laser pulse FWHM ($|y|\leq 20~\mu$m) has curvature of the opposite sign compared to the outer parts of the accelerating structure. The outer parts of the accelerating structure ($-50\leq x-x_{\rm hemi,\,center}\leq 0$, $|y|\geq 50~\mu$m) also exhibit a radius of curvature different from $R_{\rm hemi}$. This complex field structure is a consequence of the time delay between radial plasma expansion on the laser axis and away from it. Therefore, protons may experience different effective curvature of the electric field, which may steer them away from the geometrical center.

Figs.~\ref{fig:focal_shift}b–d show proton distributions in different energy bands ($\approx 2$, $\approx 5$, and $\geq 10$ MeV; measured at $t=7$ ps in the simulation) in the $(\alpha_{\rm exp},\alpha_{\rm exp}-\alpha_{\rm obs})$ space. The horizontal dashed lines in these figures correspond to $\alpha_{\rm exp}=\alpha_{\rm obs}$, i.e., protons focusing geometrically exactly to the hemisphere center. The diagonal semi-transparent dashed lines denote $\alpha_{\rm obs}=0^\circ$, corresponding to proton acceleration along the $x$ (laser) axis.

First, one may notice that the angular distributions of particles at different energies (denoted by the rainbow color) have different horizontal extent. This is expected due to the finite transverse extent of the laser pulse: its FWHM of 40 $\mu$m roughly corresponds to $|\alpha_{\rm exp}|\leq 10^\circ$. Most importantly, it is evident that proton focusing is never perfectly geometrical. Slower protons (Fig.~\ref{fig:focal_shift}b) tend to align closer to the $\alpha_{\rm exp}=\alpha_{\rm obs}$ line (geometric focusing) than other energy bands, but they deviate within the laser FWHM due to the effectively larger curvature of the accelerating field, leading to downstream focusing. The fastest particles (d) align closer to $\alpha_{\rm obs}=0$ (trajectory along x axis), which also pushes protons to focus outside the geometrical center of the hemisphere. This trend is consistent with the findings of Fig.~\ref{fig:tracking}b and suggests that the curvature of the TNSA field plays a significant role in shifting the proton focal plane away from the hemisphere center.

\subsection{The role of the hemisphere opening angle}

Besides the hemisphere radius, another important design parameter of the curved target is the opening angle, $\phi_{\rm open}$. It is defined as the full angle enclosing the curved part of the target before it transitions to the flat target holder structure. For an opening angle of 180$^\circ$, we refer to the target as a ``full'' hemisphere; for smaller angles, it is a ``partial'' hemisphere. We do not consider opening angles larger than 180$^\circ$ in this study, but note that there may be beneficial effects on proton post-acceleration and focusing in a hollow hemisphere \cite{burza2011} or in a hemisphere+cone target geometry \cite{king2023}; hollow spherical targets are also proposed to increase yield in laser-driven $\alpha$ particle sources\cite{caizergues2025}. In the context of pFI, partial hemispheres are often preferred to full hemispheres in cone-integrated target designs, because they can be coupled to a shielding cone that protects the proton source and helps guide the beam toward the cone tip\cite{king2023}. Therefore, it is crucial to understand whether they modify proton acceleration and focusing compared to a full hemisphere.

\begin{figure}
    \centering
    \includegraphics[width=\linewidth]{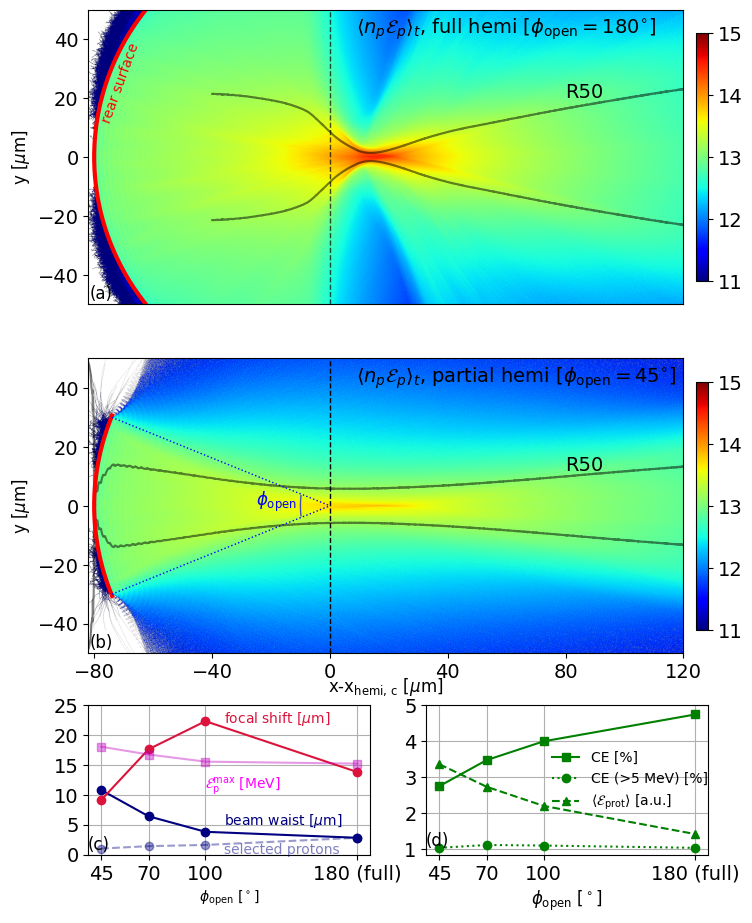}
    \caption{The role of the hemisphere opening angle, $\phi_{\rm open}$. (a) Time-integrated proton energy density map for a full hemisphere with $R_{\rm hemi}=80~\rm \mu m$. The red line denotes the target rear surface; the dashed line denotes the geometrical center of the hemisphere; solid semitransparent lines show the R50 curves (see text). (b) Time-integrated proton energy density map for a partial hemisphere with $R_{\rm hemi}=80~\rm \mu m$ and $\phi_{\rm open}=45^\circ$. (c) Proton focal shift (crimson), beam waist (navy), beam waist for the $R=80~\mu \rm m$ and $\phi_{\rm open}=180^\circ$ simulation using selected protons (semi-transparent navy dashed; see text), and proton cutoff energy (magenta) as a function of $\phi_{\rm open}$. (d) Conversion efficiency (solid squares), conversion efficiency into protons with energies $>5$ MeV (dotted circles), and average energy gain per proton (dashed triangles; arbitrary units).}
    \label{fig:partial}
\end{figure}

Fig.~\ref{fig:partial} discusses the role of the hemisphere opening angle, $\phi_{\rm open}$, in proton acceleration and focusing; the definition of $\phi_{\rm open}$ is illustrated in Fig.~\ref{fig:partial}b. First, we compare two time-integrated proton energy density maps for the full ($\phi_{\rm open}=180^\circ$) and partial ($\phi_{\rm open}=45^\circ$) hemispheres, shown in Figs.~\ref{fig:partial}a,b. Here, we again plot R50 curves to illustrate the proton focusing dynamics. It is evident that proton focusing is tighter in the case of a full hemisphere.

Inspecting the hot electron temperature distributions for the partial and full hemispheres, we found that the temperature stayed consistently more uniform in the full hemisphere case, whereas for the partial hemisphere it expanded outward from the laser FWHM throughout the simulation. Slightly higher initial hot electron temperatures (by a factor of $\approx 1.5$), along with stronger transverse temperature gradients during the early acceleration stage, resulted in a wider proton beam spot size for the partial hemisphere. Later, when the focal plasma reaches the peak compression in the full hemisphere case, the picture reverses: hot electron temperature gradients are much higher for the full hemisphere (see Figs.~\ref{fig:focalplasma},\ref{fig:focalplasmascaling}), while partial hemispheres produce much more uniform temperature and density distributions around the same location. Hence, the defocusing length (proton ``Rayleigh length'') is larger for proton beams from partial hemispheres. 

The focal shift stays positive (protons focus outside the hemisphere center) and is only weakly dependent on the opening angle within the considered range of $\phi_{\rm open}$. The proton cutoff energy is also only weakly dependent on the opening angle, with the difference between the full hemisphere and $\phi_{\rm open}=45^\circ$ around $20\%$. The higher $\mathcal{E}_{\rm p}^{\rm max}$ for the partial hemisphere is consistent with the slightly higher initial hot electron temperatures, which likely result from edge amplification effects at the sharp boundaries of the partial hemisphere. Our auxiliary simulations show that adding the flat target holder structure around the partial hemisphere produces nearly identical results.

To account for the different numbers of protons in the full and partial hemisphere simulations, we also recalculated the proton beam waist for the full hemisphere simulation while including only protons that originated within a given opening angle, $\phi_{\rm open}=45^\circ,~70^\circ,~100^\circ$ (Fig.~\ref{fig:partial}c, navy semi-transparent dashed line). These downselected protons focus to a much tighter spot than in the corresponding partial hemisphere simulations. Proton populations originating from smaller angles also focus to tighter spots in the full hemisphere simulation. This implies two effects associated with plasma converging from large angles: (I) converging plasma from large angles helps compress the focal plasma more strongly and (II) accelerated protons from large angles in the full hemisphere have sufficient energies to increase the proton spot size. Since the dependence on the hemisphere target type is stronger than the dependence on proton downselection, we conclude that the former effect dominates.

Trends in laser-to-proton conversion efficiency are presented in Fig.~\ref{fig:partial}d. The conventional conversion efficiency grows from 2.7\% to 4.7\% with increasing $\phi_{\rm open}$, as the overlap between hot electrons and protons increases for a more extended target. Conversion into fast protons (those that ended up with $\mathcal{E}_{p}>5$~MeV), however, is nearly identical for all targets, since fast protons mostly originate from within the laser FWHM. The average energy per proton is higher for smaller partial hemispheres, consistent with the total CE trend weighted by the geometric factor $\phi_{\rm open}^{-1}$.

To summarize, the hemisphere opening angle controls the proton focal spot size, while having a weak influence on the focal plane location and proton acceleration. Converging plasma from large angles on the hemisphere contributes to more efficient plasma compression, reducing the proton beam spot size.

\subsection{Laser intensity, spot size, and pulse duration scaling}\label{sec:scaling}

Finally, to study trends in proton focusing with laser parameters, we conducted a broader parametric study over a range of laser conditions. We performed a set of 2D PIC simulations with $R_{\rm hemi}=20,\,40,\,80,\,120~\mu$m and varied the laser width, $w_{\rm las}=10,\,20,\,40,\,80,\,160~\mu$m, while keeping the pulse intensity and duration fixed. For $R_{\rm hemi}=80~\mu$m, we also varied the laser intensity ($\mathcal{I}_{\rm las}=2.5\cdot10^{18}$ W/cm$^2$--$10^{20}$ W/cm$^2$) at fixed width and duration, and the laser pulse duration ($\tau_{\rm las}=40,\,300,\,1000$ fs) at fixed intensity and width.

\begin{figure}
    \centering
    \includegraphics[width=\linewidth]{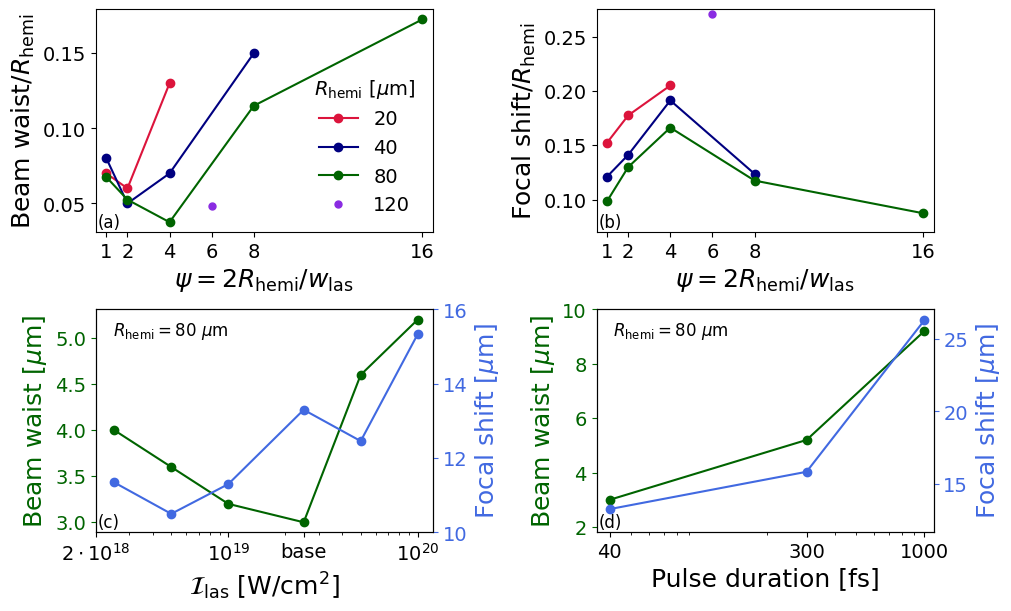}
    \caption{Scaling of proton focal characteristics with laser parameters. Dependence of (a) proton beam waist normalized to $R_{\rm hemi}$ and (b) proton focal shift normalized to $R_{\rm hemi}$ as a function of $\psi \equiv 2R_{\rm hemi}/w_{\rm las}$. Proton focal spot parameters as a function of (c) laser intensity and (d) laser pulse duration.}
    \label{fig:scaling}
\end{figure}

Figure~\ref{fig:scaling} depicts the results of our multi-parameter study of proton focusing. Subplots (a) and (b) show the proton beam waist and the shift of the proton focal plane from the geometrical center as functions of the laser width, $w_{\rm las}$, expressed in terms of the parameter $\psi \equiv 2R_{\rm hemi}/w_{\rm las}$. The results are also normalized to $R_{\rm hemi}$. The motivation for introducing these parameters is twofold. First, we aim to determine whether proton focusing dynamics is self-similar -- that is, whether proton focal characteristics for larger hemispheres can be inferred from computationally cheaper scaled-down simulations. Second, a recent numerical study of laser-driven proton focusing\cite{ospina2025} suggested that proton focal characteristics depend on the ratio of hemisphere diameter to laser spot size, identifying an optimal focusing regime with a tight focal spot and a well-collimated proton beam at $\psi \sim 6\text{--}8.5$.

For the simulations considered here, we find proton beam waists of approximately $0.05\text{--}0.15~R_{\rm hemi}$ and proton focal plane shifts around $0.1\text{--}0.25~R_{\rm hemi}$. The beam waist curve exhibits a V-shape for each $R_{\rm hemi}$ considered, indicating the existence of an optimal $\psi$ for the tightest focusing. We find the optimal $\psi$ to be somewhat smaller than that reported in Ref.~\cite{ospina2025}, likely because of differences in the laser and target parameter ranges considered in the two studies, as well as in the definition of ``optimal focusing'' used here, which does not take proton beam collimation into account. 

Remarkably, for $\psi\leq 2$, there is some evidence for a self-similar proton focusing, suggesting that for nearly uniform laser irradiation of hemisphere targets, predictions could be made based on scaled-down simulations. However, for $\psi\geq 4$, we observe a larger spread in the beam waist-$R_{\rm hemi}$ ratio, indicating that full-scale simulations are required for reliable predictions of proton focusing. The focal shifts for $\psi\leq 2$ also exhibit low variance, further suggesting self-similarity of proton focusing in this regime.

Figs.~\ref{fig:scaling}c,d demonstrate how proton focusing scales with laser intensity and pulse duration, respectively. The beam waist has a minimum near the baseline intensity of $2.6\cdot10^{19}$ W/cm$^2$, although the overall variation remains within the narrow 3--5 $\mu$m range. The focal plane tends to shift slightly inward at lower intensities, but always remains downstream of the geometrical center. For longer laser pulses, we observe broader proton beam waists located farther from the geometrical center, possibly due to emergence of magnetic field effects in picosecond pulse regime\cite{kemp2025}.

It should be noted that auxiliary simulations with plastic (CH) hemisphere targets and picosecond laser pulses suggest that proton focal plane may shift upstream of the geometrical center, with nearly flat $R50$ curves. Therefore, all trends reported here should be applied with caution to the picosecond pulse regime and to different target compositions.

\section{Discussion}\label{sec:disc}

The main findings of the manuscript may be summarized as follows:

\begin{itemize}
    \item The proton acceleration mechanism is mostly consistent with TNSA, with a minor contribution from the post-acceleration stage near the geometrical center of the hemisphere. This contribution peaks for $\psi\equiv 2R_{\rm hemi}/w_{\rm las}=1$ (uniform target illumination) and falls off for larger $\psi$.
    \item Front side curvature contributes to enhanced laser absorption (and increased CE and proton cutoff energies) via an effective oblique incidence of the pulse for $w_{\rm las}\sim R_{\rm hemi}$. Rear side curvature does not strongly affect the acceleration efficiency in our simulations.
    \item The focal plasma is found to be unmagnetized and has notable charge separation, with plasma parameters $\sim 0.01-0.1 n_{\rm cr}$ and $T_e \sim 0.5-1$ MeV, and a quasi-elliptical shape with FWHM of $\approx 6~\mu{\rm m} \times 2~\mu{\rm m}$ that weakly increases with $R_{\rm hemi}$.
    \item The proton focal spot location is downstream of the geometrical center for the bulk of the conducted simulations. The focal spot shift from the geometrical center increases with $R_{\rm hemi}$, potentially providing a useful knob for integrated hemisphere–cone targets in pFI. Both the focal spot size and the focal plane location are consistent among the different possible definitions. The virtual focal spot roughly tracks the physical focal spot but shows substantial variance because the proton trajectories are not yet fully rectilinear, complicating comparison against the experiment.
    \item Proton tracking shows the presence of fully reflecting, deflecting, and crossing trajectories, with the reflecting protons being the dominant population. The energy-dependent nature of proton focusing is evident from multiple analyses, with higher energy protons focusing further downstream. The role of the magnetic field is found to be negligible for proton focusing.
    \item The hemispherical opening angle controls the focal spot size, with smaller $\phi_{\rm open}$ leading to larger spot sizes and longer proton Rayleigh lengths. Converging plasma from larger opening angles on the hemisphere contributes to tighter plasma compression.
    \item Proton focusing is robust against variations in laser intensity, but shows linearly growing spot size and focal shift for longer pulses. Spot sizes of $\sim0.05\text{--}0.15~R_{\rm hemi}$ at locations $\sim0.1\text{--}0.25~R_{\rm hemi}$ downstream of the geometrical center are observed for a wide range of laser parameters.
    \item There is evidence of self-similar proton focusing for nearly uniform target irradiation ($\psi\leq 2$). By contrast, the regime of partial hemisphere irradiation ($\psi \geq 4$) exhibits greater variation in proton focal characteristics and likely requires full-scale simulations.
    \end{itemize}

Let us compare our findings to the experimental counterpart study \cite{griffmcmahon2026}. As the targets used in our simulations ($R_{\rm hemi}\leq 120~\mu {\rm m}$) are smaller compared to the experiments ($R_{\rm hemi}\geq 120~\mu {\rm m}$), we use the $R_{\rm hemi}=120~\mu {\rm m}$ simulation for comparison. First, because we mostly discuss Cartesian 2D simulations, our proton cutoff energies and hot electron energies are expected to be elevated compared to the real 3D system\cite{xiao2018}. We obtain a cutoff energy of 18 MeV, which is $\sim$3 times larger than the value observed in the experiment. We find a similar proton beam waist (6 $\mu$m, from Fig.~\ref{fig:R50waistshift}b vs the experimental 9$\pm$3 $\mu{\rm m}$). However, the absolute focal locations are different: our simulations show proton focusing exclusively outside the hemisphere, whereas the experiment infers focusing inside the hemisphere. Our laser parameter scan suggests that decreased laser intensity and/or a smaller laser spot may contribute to shifting the \textit{virtual} proton focal plane upstream of the geometrical center. Therefore, one potential hypothesis that may explain the experimental observations is that the effects of hot electron propagation through the target are not included in our simulations and, to account for them, we may need to adjust the laser pulse parameters to generate a ``correct'' hot electron cloud on the rear side of the target. Fast electron propagation through the target typically leads to a decrease in electron temperature and an increase in the electron beam waist (although pinching via self-generated magnetic fields is possible\cite{bell2003,kar2009}). In collisionless PIC simulations, these effects would correspond to a lower effective laser intensity and a different (likely, larger) effective laser waist.

The detailed comparison against the broader published literature is complicated, as the laser and target parameters used in our study are qualitatively different from other studies \cite{kar2011,bartal2012,offermann2011,chen2012,snavely2007}. Some trends, such as tighter focal spots for higher energy protons \cite{kar2011,bartal2012}, proton focusing at or downstream of the geometrical center \cite{kar2011,chen2012}, and energy-dependent focusing \cite{chen2012}, agree with our simulations, whereas others, such as a focal spot location inside the hemisphere (or within the curvature radius) and the absolute values of the proton spot size \cite{offermann2011,bartal2012}, disagree. We note that our auxiliary simulations with longer pulses (OMEGA EP-like) recover some of these trends, yielding wider focal spots well inside the hemisphere radius.

Analytical models, such as the one developed in Refs.~\cite{dorozhkina1998,bellei2012}, may be useful for predicting the long-term evolution of the proton beam, provided their assumptions are satisfied. We tested the assumptions underlying the proposed self-similar solution: quasi-neutrality, electrostatic interaction, and a specific ansatz relating the spatial and velocity distributions of the plasma. We found that the quasi-neutrality assumption is marginally satisfied, and the electrostatic interaction assumption is fair. We also performed statistical tests to assess whether the focal plasma distributions observed in the simulations are consistent with a Gaussian distribution in space and a bi-Maxwellian distribution in velocity space for both electrons and protons. We found that this is not the case for either species. We therefore conclude that a more sophisticated analytical model is needed for estimating long-term proton beam dynamics.

One of the parameters in our simulations is the contamination layer conditions. For the bulk of our simulations, we considered a purely H contamination layer deposited only on the rear side of the target, with the density exponentially falling off over a 50~nm thickness with an e-folding length of 10~nm. While the measured thickness of contamination layers is typically on the order of a few nm \cite{lecz2020}, it is common practice to use thicker layers in simulations \cite{sgattoni2012,park2020}. We note that the proton areal density in our baseline simulations is $\approx 4.4\cdot 10^{16}$ cm$^{-2}$, consistent with experimental measurements (few$\times 10^{16}$ cm$^{-2}$; see Ref.~\cite{hoffmeister2013}). Auxiliary simulations considered the role of front side H, the addition of oxygen ions on the rear side, and a reduced peak proton density of the layer in separate runs. We found the first two to play only a minor role in proton acceleration and focusing. The initial proton layer density, however, may control the transition from purely ballistic motion at low proton densities (0.4$n_{\rm cr}$ peak proton density, 100 times smaller than the baseline simulation) to collective plasma behavior for the baseline simulation parameters. As a result, the focal shift and spot size may change by up to a factor of two, shifting the focal spot closer to the geometrical center of the hemisphere (though still downstream of it) and increasing the proton beam waist in the low density simulation. Because our baseline simulations demonstrate collective plasma behavior and are in reasonable agreement with the published contamination layer measurements, collective behavior of the electron–proton plasma may be expected under the experimental conditions of Ref.~\cite{griffmcmahon2026} as well.

Two experimental uncertainties that could contribute to variations in proton focusing are finite laser contrast and laser pointing stability. Laser prepulses may alter the plasma conditions on the front side of the target, modifying the interaction of the driver pulse with the target. Available measurements for the CSU ALEPH laser indicate a contrast level of $10^{8}$ at 500 ps for the fundamental harmonic. We conducted simulations with the radiation hydrodynamic code FLASH\cite{fryxell2000,tzeferacos2015} to estimate preplasma conditions for a range of prepulse durations and intensities for aluminum targets and CSU ALEPH-inspired prepulses. Considering laser contrasts of $10^{8}$ and $10^{6}$ and prepulse durations of 100--500 ps, we obtained preplasma profiles, inserted them into 2D EPOCH simulations, and studied proton acceleration and focusing under various preplasma conditions. We find that the laser-to-proton CE increases for more energetic prepulses, as commonly reported in the literature\cite{gizzi2021}, while the focal characteristics remain largely insensitive to laser contrast.

Laser pointing errors could lead to a mismatch between the hemisphere axis of symmetry and the laser axis, affecting proton focusing. Auxiliary simulations with shifted laser axis showed that the location of the proton focal spot, defined as the peak of the time-integrated proton energy density, is rather insensitive to laser axis shifts on the order of the laser spot size, $\delta y \sim w_{\rm las}$. At the same time, we identified two trends in the simulations that could be compared against the experiment. First, a large laser axis shift could result in effectively oblique laser incidence on the hemispherical target, which may enhance coupling and increase proton cutoff energies, similarly to what we observed in the $w_{\rm las}\sim R_{\rm hemi}$ regime. Second, we find that fast protons tend to propagate along the axis connecting the laser impact point on the target and the geometrical center of the hemisphere, while lower energy protons remain more isotropic in their angular distribution. This could manifest as energy-dependent shifts in the proton fluence spot location and could be tested experimentally.

Ref.\cite{higginson2026} explores a complementary parameter space, focusing on Trident laser conditions relevant to picosecond pulse-driven proton acceleration and focusing. Ref.~\cite{kemp2025}, as well as our auxiliary simulations, show that proton focusing can differ for longer, higher energy pulses, as the contribution of electron currents and the associated magnetic fields becomes more noticeable. Still, the main trends, such as a proton focal plane location that scales linearly with $R_{\rm hemi}$ and lies downstream of the curvature radius, the energy dependence of proton focusing, and the ranges of proton focal shift ($\sim 0.1\text{--}0.25~R_{\rm hemi}$) and spot size ($\sim 0.05\text{--}0.15~R_{\rm hemi}$), are broadly consistent. The main discrepancy appears to be the dependence on laser intensity. Ref.~\cite{higginson2026} finds that decreasing laser intensity moves the focal plane upstream ($z_{\rm foc}\propto \mathcal{I}_{L}^{-1/4}$), whereas our scans show the opposite trend. We hypothesize that, in addition to the different pulse duration regimes, this difference may be related to the hemisphere opening angles. The opening angle was shown to modify the spot size in our simulations and in Ref.~\cite{higginson2026}; most of our results are for a full hemisphere, while Ref.~\cite{higginson2026} mainly considers $\phi_{\rm open}=54^\circ$. The general agreement between two different numerical approaches (Cartesian 2D full PIC in this manuscript vs cylindrical 2D hybrid PIC in Ref.~\cite{higginson2026}) is encouraging and motivates a more detailed code comparison study.

To conclude, this manuscript studied laser-driven proton acceleration and focusing from concave targets. TNSA was found to dominate proton acceleration, with proton beam waist and focal plane shift scaling approximately linearly with the hemisphere radius. We also observed chromatic behavior of hemispherical targets, with different proton energy bands focusing at different locations. A broader parametric scan reveals indications of self-similar proton focusing in the regime of nearly uniform target irradiation ($\psi \equiv 2R_{\rm hemi}/w_{\rm las}\leq 2$), raising the possibility that parameters of interest can be inferred from scaled-down simulations with smaller target sizes. In contrast, this self-similarity breaks down for partial irradiation ($\psi \geq 4$). Finally, while the predicted focusing characteristics are broadly consistent with results obtained in the picosecond regime\cite{kemp2025,higginson2026}, caution is required when extrapolating these findings to significantly different laser or target conditions.

\section*{Acknowledgements}
The research was conducted under the Laboratory Directed Research and Development (LDRD) Program at Princeton Plasma Physics Laboratory, a national laboratory operated by Princeton University for the U.S. Department of Energy under Prime Contract No. DE-AC02-09CH11466. V.O-B. and X.V. acknowledge the support from Focused Energy as part of its Inertial Fusion Energy Research and Development program. This research was supported in part by grant NSF PHY-2309135 to the Kavli Institute for Theoretical Physics (KITP). The simulations presented in this article were performed on computational resources managed and supported by Princeton Research Computing at Princeton University. This research used resources of the National Energy Research Scientific Computing Center (NERSC), a Department of Energy User Facility using NERSC awards FES-ERCAP0031176 and FES-ERCAP0035525.

\bibliographystyle{apsrev4-1}
\bibliography{ion_acceleration}

\appendix

\section{Effects of dimensionality}\label{sec:3D}

\begin{figure}
    \centering
    \includegraphics[width=\linewidth]{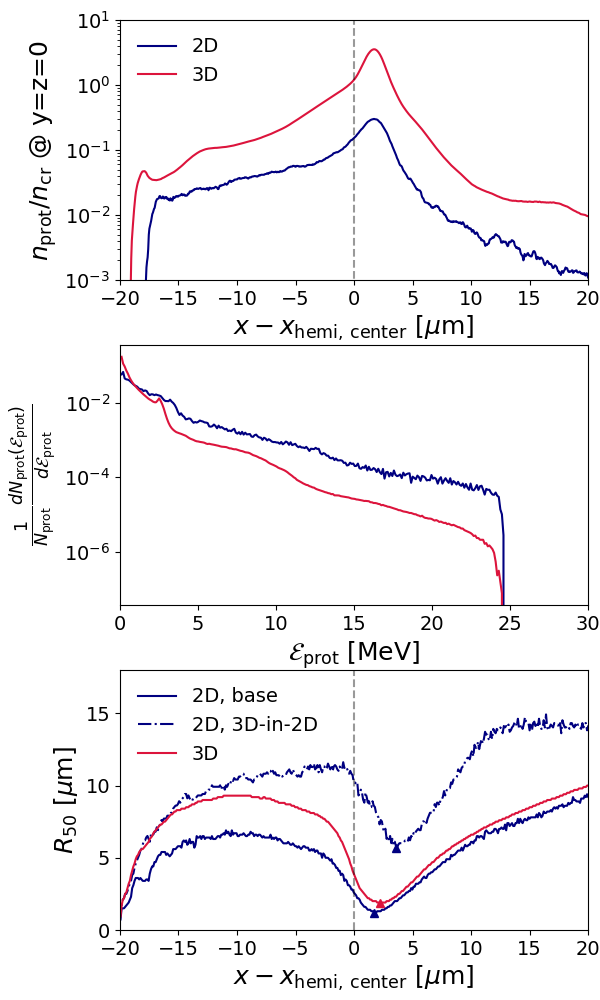}
    \caption{Comparison of 2D and 3D simulations of laser-driven proton focusing for $R_{\rm hemi}=20~\mu$m. (a) 1D cut at y=z=0 of proton density for 2D (navy) and 3D (crimson) around the hemisphere center (dashed gray line). (b) Proton energy spectra normalized to the total number of particles. (c) R50 curves from the baseline 2D simulation (solid navy), calculation with 2D simulation assuming cylindrical symmetry (``3D-in-2D'', dot-dashed navy), and 3D simulation. Focal points are denoted with upward facing triangles of respective color.}
    \label{fig:2d_vs_3d}
\end{figure}

To assess the role of dimensional effects, we conducted a 3D simulation of laser-driven proton focusing from a hemisphere target and compared it with the corresponding 2D simulation. To make the 3D simulation computationally feasible, we modified the problem parameters slightly and performed a matching 2D simulation to ensure a fair comparison.

For the 2D simulation, we used lower grid ($N_{\rm grid}=20$) and particle resolutions ($N_{\rm ppc}=10$ for electrons, $N_{\rm ppc}=2$ for gold, and $N_{\rm ppc}=100$ for protons) as compared to the primary set of simulations, reduced the box size to $100~\mu\rm m\times 60~\mu\rm m$, and considered a hemisphere with $R_{\rm hemi}=20~\mu \rm m$ and thickness of $2\,\mu \rm m$, with contamination layer and preplasma defined similarly to those in the main text. Such changes only slightly modified the resulting R50 curve and proton focal characteristics. The 3D simulation used the same initial conditions, with the density profiles ``rotated'' around the $y=z=0$ axis. Both simulations were run for 2 ps.

Figure~\ref{fig:setup}b sketches out the results of a 3D simulation. It depicts the initial location of a hemisphere target (yellow colored volume), laser beam (crimson), and the time-integrated proton energy density map. It is notable that the proton beam distribution is not rotationally symmetric around the $y=z=0$ axis, suggesting a potential effect of linear laser polarization (laser $E$ field is in x-y plane) on proton focusing. A similar effect is observed when varying the angle of linear laser polarization (in-plane or out-of-plane laser $E$ field) in 2D PIC simulations. Detailed investigation of this phenomenon is beyond the scope of this paper.

Figure~\ref{fig:2d_vs_3d} presents the main simulation results. Fig.\ref{fig:2d_vs_3d}a compares the proton density on the laser axis ($y=0$ in 2D, navy; $y=z=0$ in 3D, crimson) at $t=1$ ps. Proton compression is stronger in 3D, as expected, since 3D simulations generally exhibit weaker sheath fields and lower hot electron temperatures than 2D simulations, while geometric convergence also favors stronger on-axis compression in 3D compared to 2D \cite{xiao2018}. The proton energy spectra shown in Fig.\ref{fig:2d_vs_3d}b are quite similar. It should be noted, however, that at earlier times the cutoff energies in the 2D case exceed those in 3D by about 20\%, suggesting that the close agreement of cutoff energies at $t=1$ ps is merely coincidental. The laser-to-proton conversion efficiencies are 4.2\% and 1.8\% in 2D and 3D, respectively, consistent with the conversion efficiency being overstated in 2D by a factor of a few\cite{xiao2018}. Finally, the R50 curves are shown in Fig.\ref{fig:2d_vs_3d}c. We adopted two different approaches in 2D: a baseline approach, in which R50 is calculated by integrating the proton energy density along the vertical direction for each $x$ location, and a ``3D-in-2D'' approach, in which the resulting 2D proton energy density distribution is interpreted as a cut through an axisymmetric 3D simulation. The resulting focal points are denoted by upward facing triangles of the corresponding colors.

We find that the proton focal characteristics obtained with these methods are qualitatively similar. The focal plane locations are comparable in all three cases. Comparing the ``3D-in-2D'' and 3D metrics, we find that protons are focused to a narrower spot in 3D, which is again consistent with the reasoning above -- lower electron temperatures and different geometrical factors of plasma convergence.

We therefore conclude that, while several well-known dimensionality effects are present in our simulations, the qualitative proton focusing behavior is similar in 2D and 3D, thereby justifying the use of the 2D approach in the main body of the manuscript.

\end{document}